\documentclass[12pt]{article}
\usepackage{amscd,amsmath,amssymb,amstext,amsthm,exscale,latexsym}
\usepackage{graphicx,floatflt}
\newcommand {\al}   {\alpha}       
\newcommand {\g }   {\gamma}       \newcommand {\G }  {\Gamma}
\newcommand {\dl}   {\delta}       \newcommand {\e }  {\epsilon}
\newcommand {\z }   {\zeta}        
\newcommand {\ve}   {\varepsilon}  
\newcommand {\lm}   {\lambda}      
          
\newcommand {\s }   {\sigma}       
         
\newcommand {\vf }  {\varphi}      
         
\newcommand {\Lm}   {\Lambda}      
       \newcommand {\Ph}  {\Phi}
\newcommand {\pl}   {\partial}     \newcommand {\nb}  {\nabla}
\newcommand   {\sign}{{\sf\,sign\,}}
\newcommand   {\const}{{\sf\,const}}
\newcommand   {\diag}{{\sf\,diag\,}}
\newcommand   {\arctanh}{{\sf\,arcth\,}}
\newcommand   {\ex}{{\sf\,e}}
\newcommand {\MG}  {{\mathbb G}}
\newcommand {\MM}  {{\mathbb M}}
\newcommand {\MS}  {{\mathbb S}}
\newcommand {\MO}  {{\mathbb O}}
\newcommand {\MU}  {{\mathbb U}}
\newcommand {\MR}  {{\mathbb R}}

\begin{document}
\title     {Inside the BTZ black hole}
\author    {G. de Berredo-Peixoto
            \thanks{E-mail: guilherme@fisica.ufjf.br}\\ \\
            \sl Departamento de Fisica, Universidade Federal de Juiz de Fora,\\
            \sl Juiz de Fora. CEP 36036--330, MG, Brazil\\ \\
            M. O. Katanaev
            \thanks{E-mail: katanaev@mi.ras.ru}\\ \\
            \sl Steklov Mathematical Institute,\\
            \sl Gubkin St. 8, Moscow, 119991, Russia}
\date      {22 November 2006}
\maketitle
\begin{abstract}
  We consider static circularly symmetric solution of three-dimensional
  Einstein's equations with negative cosmological constant (the BTZ black
  hole). The case of zero cosmological constant corresponding to the
  interior region of a black hole is analyzed in detail. We prove that
  the maximally extended BTZ solution with zero cosmological constant
  coincides with flat three-dimensional Minkowskian space-time without
  any singularity and horizons. The Euclidean version of this solution
  is shown to have physical interpretation in the geometric theory of
  defects in solids describing combined wedge and screw dislocations.
\end{abstract}
\vskip10mm
\section{Introduction}
Three-dimensional static circularly symmetric solution of three-dimensional
Einstein's equations (the BTZ black hole \cite{BaTeZa92}) attracted much
interest last years, and is believed to be a good relatively simple
laboratory for analyzing general aspects of black hole physics. The
BTZ solution is the most general black hole solution in three dimensions,
which is guaranteed by the three-dimensional version of Birkhoff's
theorem \cite{AyMaZa04}. It has very interesting classical and quantum
properties (for review see, i.e. \cite{Carlip98}).

Global structure of the BTZ solution was analyzed in \cite{BaHeTeZa93,Steif96}.
In these articles the black hole space-time was considered as the quotient
space of the anti-de Sitter space by the action of the discrete transformation
group. This approach deserves a deeper analysis because the transformation
group has fixed points, and therefore the quotient space is not a manifold.
The situation is unclear especially in the interior region of the black hole.
To clarify the global structure of the interior region, we consider the BTZ
solution for zero cosmological constant. In this simple case, all coordinate
transformations can be written explicitly, and behavior of geodesics is
analyzed. We prove that the maximally extended (along geodesics) BTZ solution
for zero cosmological constant coincides with flat Minkowskian space-time
without any singularity and horizons for infinite range of the angle coordinate
$\vf$. The BTZ solution in the original coordinates covers only one half of the
Minkowskian space-time, and the two planes corresponding to $r=0$ are just
coordinate singularities. Two copies of the BTZ solution cover the whole
Minkowskian space-time and are smoothly glued at $r=0$.

If we make the angle identification $\vf\sim\vf+2\pi$ than four cones with
the same vertex appear in the interior region of the BTZ black hole located
at the inner horizon $r_-$.

Next we analyze the Euclidean version of the BTZ solution. In this case
the solution has physical interpretation in solid state physics.
In the geometric theory of defects developed in
\cite{KatVol92,KatVol99,Katana03,Katana04} (for review see \cite{Katana05})
it depends on the Poisson ratio and describes combined wedge and screw
dislocations.
\section{The BTZ solution}
We consider a three-dimensional manifold $\MM$ with local coordinates
$x^\mu$, $\mu=0,1,2$. Suppose it is endowed with a
Lorentzian signature metric $g_{\mu\nu}(x)$, $\sign g_{\mu\nu}=(+--)$,
which satisfies Einstein's equations
\begin{equation}
  R_{\mu\nu}=\Lm g_{\mu\nu}
\end{equation}
with a cosmological constant $\Lm=-2/l^2$. In three dimensions the full curvature
tensor is defined by its Ricci tensor
\begin{equation*}
  R_{\mu\nu\rho\s}=-\ve_{\mu\nu\lm}\ve_{\rho\s\z}R^{\lm\z},~~~~
  R_{\mu\nu}=R_{\mu\rho\nu}{}^\rho
\end{equation*}
where $\e_{\mu\nu\lm}$ is the totally antisymmetric tensor,
$\e_{012}=\sqrt{\det g_{\mu\nu}}$. Therefore, any smooth solution of Einstein's
equations on a manifold (by solution we mean a pair $(\MM,g)$) is a space-time
of constant curvature. The universal covering spaces for positive, zero, and
negative cosmological constant are respectively de Sitter, Euclidean, and
anti de Sitter spaces. All other smooth solutions $(\MM,g)$ are obtained
from these solutions by an action of an isometry transformation group which
acts freely and properly discontinuous (see, i.e.\ \cite{Wolf72}). Afterwards,
a solution will be isometric to some fundamental domain of de Sitter,
Euclidean, or anti de Sitter space with properly identified boundaries.
These solutions are considered as known ones though explicit action of a
transformation group may be quite complicated. All these solutions are smooth,
have no singularities and horizons, and therefore do not describe black holes.

Theory becomes much richer if we admit the existence of singularities at
points, lines, or surfaces in $\MM$. The famous BTZ solution is \cite{BaTeZa92}
\begin{equation}                                        \label{ebtzso}
\begin{split}
  ds^2&=\left(-M+\frac{J^2}{4r^2}+\frac{r^2}{l^2}\right)dt^2
  -\frac{dr^2}{\left(-M+\frac{J^2}{4r^2}+\frac{r^2}{l^2}\right)}
  -r^2\left(d\vf-\frac J{2r^2}dt\right)^2,
\\
  &=\left(-M+\frac{r^2}{l^2}\right)dt^2
  -\frac{dr^2}{\left(-M+\frac{J^2}{4r^2}+\frac{r^2}{l^2}\right)}
  -r^2d\vf^2+Jdtd\vf,
\end{split}
\end{equation}
where $M$ and $J$ are two integration constants, having physical
interpretation of the mass and angular momentum of the black hole.
We shall see later that the inner horizon $r_-$ becomes a line in $\MM$
with four cones at each point.
This solution is supposed to be written in cylindrical coordinate system
\begin{equation}                                         \label{eracod}
  t\in(-\infty,\infty),~~r\in(0,\infty),~~\vf\in(0,2\pi).
\end{equation}

Metric (\ref{ebtzso}) has two commuting Killing vector fields: $K_1=\pl_t$ and
$K_2=\pl_\vf$.

Outer $r_+$ and inner $r_-$ horizons of the BTZ solution
\begin{equation*}
  r^2_\pm=\frac{Ml^2}2\left(1\pm\sqrt{1-\frac{J^2}{M^2l^2}}\right)
\end{equation*}
are defined by two positive zeroes of the function
\begin{equation*}
  N(r)=-M+\frac{J^2}{4r^2}+\frac{r^2}{l^2}=0.
\end{equation*}
Here we assume that $|J|<Ml$.

We make four comments on the form of this solution.

1) The space-time $(M,g)$ with metric (\ref{ebtzso}), (\ref{eracod}) is not
locally a constant curvature space, because any point lying in the singularity
$r=r_-$ do not have a neighborhood of constant curvature. Here we consider
singular points $(t,r=r_-,\vf)$ belonging to $\MM$. In fact, we shall see that
points $r=r_-$ are not points of a manifold, and do not have neighborhoods
diffeomorphic to a ball at all.

2) The range of $\vf$ is determined by its interpretation as an angle in the
region $r\to\infty$ where the space-time becomes asymptotically anti-de Sitter.
The mass is supposed to be positive, because otherwise there is no horizon.
The case $M<0$ and $\Lm>0$, corresponding to de Sitter asymptotic (which has a
horizon) is not considered because $\vf$ can not be interpreted as the angle
coordinate in this case. The sign of $J$ corresponds to left and right rotations
and does not contribute to the global structure of the solution. Thus, range
of coordinates (\ref{eracod}) and the signs of the cosmological constant and
the mass are uniquely determined by two requirements: (i) $\vf$ is an angle
in the asymptotic region $r\to\infty$ and (ii) the solution has at least
one horizon.

3) For zero cosmological constant $\Lm=0$ and angular momentum $J=0$ the metric is
\begin{equation*}
  ds^2=-Mdt^2+\frac{dr^2}M-r^2d\vf^2.
\end{equation*}
For positive mass $M>0$, it has no conical singularity at $r=0$ because now $r$
and $\vf$ are coordinates on the two-dimensional Minkowskian space-time of
signature $(+-)$ but not on the Euclidean plane. This is in contrast with
a common belief that static point particles in three-dimensional gravity are
described by conical singularities, distributed on a space-like section of $\MM$
\cite{Starus63,Clemen76,DeJatH84}.

4) There are four regions of $r$ where coordinate lines $t,r$, and $\vf$ have
different types of tangent vectors. Coordinate lines $t$ and $r$ may be either
timelike or spacelike, depending on range of $r$ which has three distinguished
points: $r_+,r_-$, and $r_3=Ml^2$. The coordinate line $\vf$ is always spacelike.
We summarize different properties of coordinates in the Table, where plus
and minus signs denote respectively timelike and spacelike character of
coordinates.
\begin{table}[h]
\begin{center}
    \begin{tabular}{|c|c|c|c|}                                  \hline
     & $t$ & $r$ & $\vf$            \\   \hline
     $r_3<r<\infty$ & $+$ & $-$ & $-$  \\   \hline
     $r_+<r<r_3$    & $-$ & $-$ & $-$  \\   \hline
     $r_-<r<r_+$    & $-$ & $+$ & $-$  \\   \hline
     $0<r<r_-$      & $-$ & $-$ & $-$  \\   \hline
    \end{tabular}
    \caption{\label{tdegeo} Timelike ``$+$'' and spacelike ``$-$'' character
    of coordinate lines in different regions of $r$.}
\end{center}
\end{table}
We see that the coordinate $t$ is timelike only for large $r>r_3$. In two
regions, $r_+<r<r_3$ and $0<r<r_-$, all coordinates are spacelike.

Global structure of solution (\ref{ebtzso}) was described in \cite{BaHeTeZa93}
in terms of the quotient space of the anti-de Sitter space, and
the Carter--Penrose diagrams were drawn for the metric
\begin{equation}                                             \label{etomse}
  dl^2=Ndt^2-N^{-1}dr^2.
\end{equation}
This description deserves a deeper analysis especially at the
vicinity of the singularity for two reasons. First, the transformation
group used to obtain the quotient of the anti-de Sitter space in
\cite{BaHeTeZa93} does not act freely because it has fixed points
\cite{Steif96}. Second, the coordinate lines have different character,
and the induced metric on sections of $\MM$ corresponding to constant angle
$\vf=\const$ differs essentially from (\ref{etomse}). Therefore we can not
say that the global solution is topologically the product of the
Carter--Penrose diagram on a circle. In the next section we give a different
description of the global structure in terms of the product of real line
$t'\in\MR$ on the corresponding Carter--Penrose diagram.
\section{The interior region of the BTZ black hole}
By global solution we mean a pair $(\MM,g)$ where (i) the metric satisfies
Einstein's equations and (ii) any extremal (geodesic) either can be
continued in both directions to an infinite value of the canonical parameter
or it ends up at a singular point at a finite value. This solution is also
called maximally extended. To construct a maximally extended solution for
BTZ metric (\ref{ebtzso}) we shall use a conformal block method described
in \cite{Katana00A} which was developed for two-dimensional metrics
having one Killing vector field.

To analyze global structure of the interior region of the BTZ black hole,
we assume that cosmological constant is zero, $\Lm=0$ or $l\to\infty$.
This assumption simplifies essentially formulae, which now may be written
explicitly. Moreover, in this case the solution has direct interpretation
in the geometric theory of defects describing a linear dislocation which
is a combination of screw and wedge dislocations (see section \ref{sscwed}).

\begin{figure}[h,b,t]
\hfill\includegraphics[height=60mm]{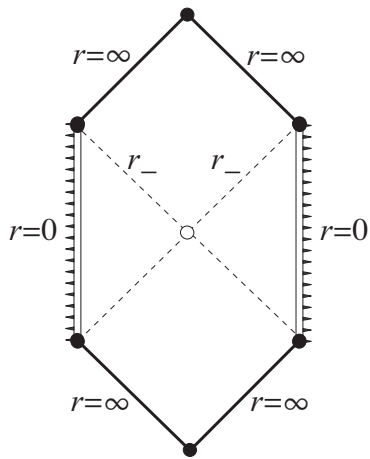}
\hfill {}
\\
\centering \caption{The Carter--Penrose diagram corresponding to a surface
with metric (\ref{etomse}), (\ref{ecoffi}). Solid lines denote complete
null infinity. The timelike boundary $r=0$ is incomplete, and the
two-dimensional curvature is singular here. Filled circles denote complete
space and time infinities. \label{fcpwro}}
\end{figure}

For later comparison, we draw the Carter--Penrose diagram for two-dimensional
metric (\ref{etomse}) with
\begin{equation}                                            \label{ecoffi}
  N(r)=-\al^2+\frac{c^2}{r^2},
\end{equation}
where for simplicity we introduced new notations $M=\al^2$ and $c=J/2$.
In the geometric theory of defects, $\al=1+\theta$, where $\theta$ is the
deficit angle of the wedge dislocation, and $c=b/2\pi$, where $b$ is the
Burgers vector of the screw dislocation. The function (\ref{ecoffi})
becomes a conformal factor in coordinates $t,r'$ where new radial
coordinate $r'$ is determined by the ordinary differential equation
\begin{equation*}
  \frac{dr}{dr'}=N(r).
\end{equation*}
In these coordinates, the global structure is easily analyzed \cite{Katana00A},
and the Carter--Penrose diagram for the surface with metric (\ref{etomse}),
(\ref{ecoffi}) is shown in Fig.~\ref{fcpwro}
It has one horizon at $r_-$. In the limit $l\to\infty$ the outer horizon
$r_+$ moves to infinity and the boundary $r=\infty$ becomes geodesically
complete which is shown by solid thick lines. At the boundaries $r=0$,
the two dimensional curvature for metric (\ref{etomse}), (\ref{ecoffi})
is singular. Filled circles denote complete future, past and right, left
``infinities''. A circle in the center denotes incomplete vertex of
conformal blocks. Unfortunately, the induced metric on sections of $\MM$
corresponding to $\vf=\const$
\begin{equation*}
  dl^2=-\al^2dt^2-\frac{dr^2}{-\al^2+\frac{c^2}{r^2}}
\end{equation*}
is quite different from (\ref{etomse}), (\ref{ecoffi}). This metric is
degenerate at the horizon $r_-=c/\al$, where it changes signature.
Therefore, we are not able to draw at least one Carter--Penrose diagram
for it.

\begin{figure}[h,b,t]
\hfill\includegraphics[height=60mm]{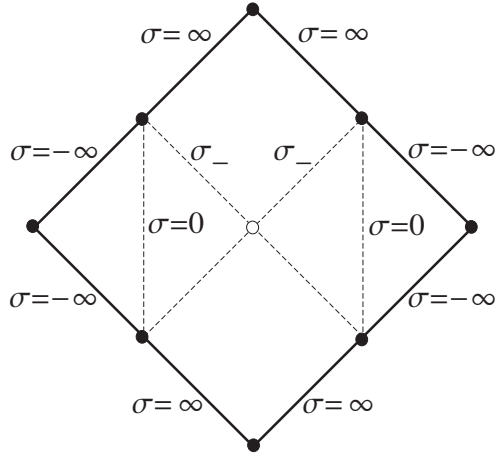}
\hfill {}
\\
\centering \caption{The Carter--Penrose diagram corresponding to the
surface $\MU$ with metric (\ref{etwmes}). It represents two-dimensional
flat Minkowskian plane in $\s,\vf$ coordinates. The boundary is complete.
Two copies of the BTZ solution are smoothly glued together along timelike
dashed lines $\s=0$. \label{fcpfou}}
\end{figure}

To avoid this difficulty, we draw the Carter--Penrose diagram for
sections of constant ``time'' which is a spacelike coordinate inside
the BTZ black hole. First, we diagonalize the metric on $\MM$
\begin{equation}                                       \label{ebtzma}
  ds^2=-\al^2dt^2-\frac{dr^2}{-\al^2+\frac{c^2}{r^2}}-r^2d\vf^2+2cd\vf dt.
\end{equation}
Performing the linear nondegenerate coordinate transformation
\begin{equation}                                             \label{enewti}
  t=t'+\frac c{\al^2}\vf,
\end{equation}
and keeping coordinates $r$ and $\vf$ untouched, we obtain
\begin{equation}                                             \label{emetrs}
  ds^2=-\al^2dt^{\prime2}-\frac{dr^2}{-\al^2+\frac{c^2}{r^2}}
  -\left(r^2-\frac{c^2}{\al^2}\right)d\vf^2.
\end{equation}
Now the three-dimensional space-time can be represented as a product
$\MM=\MR\times\MU$ where $t'\in\MR$ and $(r,\vf)\in\MU$. The two-dimensional
surface $\MU$ (sections of $\MM$ corresponding to $t'=\const$) possesses the
induced metric of Lorentzian signature. At this point we assume that
$\vf\in\MR$ making the identification $\vf\sim\vf+2\pi$ afterwards. Changing
the radial coordinate,
\begin{equation}                                             \label{echrad}
  r=\sqrt{2\al\s},~~~~\s>0~~\text{for}~~\al>0,
\end{equation}
the induced metric on $\MU$ becomes
\begin{equation}                                             \label{etwmes}
  dl^2=-\frac{\al^2 d\s^2}{c^2-2\al^3\s}+\frac{c^2-2\al^3\s}{\al^2}d\vf^2.
\end{equation}
The conformal factor for the induced metric $N(\s)=(c^2-2\al^3\s)/\al^2$ is
a linear function of $\s$ and therefore describes a surface of zero
two-dimensional curvature. It has one horizon at $\s_-=c^2/2\al^3$ corresponding
to inner horizon $r_-$. The maximally extended surfaces $(\s,\vf)\in\MU$ is
represented by the Carter--Penrose diagram in Fig.~\ref{fcpfou}.
Here the maximal extension requires the infinite range of coordinates
$\s\in\MR$ and $\vf\in\MR$. To consolidate this range of coordinates
with the original radial coordinate (\ref{echrad}) we assume that
$\s>0$ for $\al>0$ and $\s<0$ for $\al<0$, and make the identification
$(\s,\al)\sim(-\s,-\al)$. We see that after the identification we must
consider two-dimensional surfaces $r=0$ for $\MM$ as the boundary.
\subsection{Geodesics}
The behavior of geodesics for the BTZ solution was considered in
\cite{FaGaSe93,CrMaPe94}. In the case of zero cosmological constant
equations for geodesics are greatly simplified, and a general solution
can be written in elementary functions and analyzed in detail.

To understand the nature of the singularity at $r_-$ we analyze the
behavior of geodesics
$x^\mu(\tau)=\left(\vphantom{\frac ab}t(\tau),r(\tau),\vf(\tau)\right)$
for metric (\ref{ebtzma}) in this section.
The only nonzero Christoffel's symbols are
\begin{equation*}
\begin{aligned}
  \G_{11}{}^1&=-\frac{c^2}{r(\al^2r^2-c^2)}, &\qquad
  \G_{12}{}^0=\G_{21}{}^0&=\frac{cr}{\al^2r^2-c^2},
\\[2mm]
  \G_{22}{}^1&=\frac{\al^2r^2-c^2}r, & \qquad
  \G_{12}{}^2=\G_{21}{}^2&=\frac{\al^2r}{\al^2r^2-c^2}.
\end{aligned}
\end{equation*}
The equations for geodesics
\begin{equation}                                        \label{egebtz}
  \ddot x^\mu=-\G_{\nu\rho}{}^\mu\dot x^\nu\dot x^\rho,
\end{equation}
where dots denote differentiation with respect to the canonical parameter
$\tau$, become
\begin{equation}                                        \label{egebtn}
\begin{split}
  \ddot t&=-2\frac{cr}{\al^2 r^2-c^2}\dot r\dot \vf,
\\
  \ddot r&=\frac{c^2}{r(\al^2r^2-c^2)}\dot r^2-\frac{\al^2r^2-c^2}r\dot\vf^2,
\\
  \ddot \vf&=-2\frac{\al^2r}{\al^2r^2-c^2}\dot r\dot\vf.
\end{split}
\end{equation}
These equations can be integrated in elementary functions. First, we
note that any system of equations for geodesics has a conserved quantity:
the length of a tangent vector $C_0=g_{\mu\nu}\dot x^\mu\dot x^\nu$.
It corresponds to conservation of energy of a point particle which, by
assumption, moves along a geodesic line
\begin{equation}                                        \label{econen}
  C_0=-\al^2\dot t^2+\frac{r^2}{\al^2 r^2-c^2}\dot r^2-r^2\dot\vf^2+2c\dot t\dot\vf.
\end{equation}
Another two conservations laws $C_{1,2}=g_{\mu\nu}\dot x^\mu K_{1,2}^\nu$
correspond to the symmetry of the metric, generated by two Killing vector
fields: $K_1=\pl_t$ and $K_2=\pl_\vf$. They describe respectively
conservation of momenta and angular momenta
\begin{align}                                           \label{econfi}
  C_1&=-\al^2\dot t+c\dot\vf,
\\                                                      \label{econse}
  C_2&=~~c\dot t-r^2\dot\vf.
\end{align}

Conservation laws (\ref{econen})--(\ref{econse}) can be easily solved
with respect to the first derivatives
\begin{equation}                                        \label{efilod}
\begin{split}
  \dot t&=-\frac{r^2C_1+cC_2}{\al^2r^2-c^2},
\\
  \dot r^2&=\frac{\al^2r^2-c^2}{r^2}C_0+\frac{r^2C_1^2+2cC_1C_2+\al^2C_2^2}{r^2},
\\
  \dot\vf&=-\frac{cC_1+\al^2C_2}{\al^2r^2-c^2}.
\end{split}
\end{equation}
These equations are compared with their Euclidean counterpart in section
\ref{sgeoeu}. We show that the connected Lorentzian manifold breaks into
disconnected pieces along horizons in the Euclidean case.

Equations for geodesics (\ref{efilod}) can be integrated explicitly.
To make clear further integration we perform the transformation to
Cartesian coordinates.
\subsection{Transformation to Cartesian coordinates}
\begin{figure}[h,b,t]
\hfill\includegraphics[height=60mm]{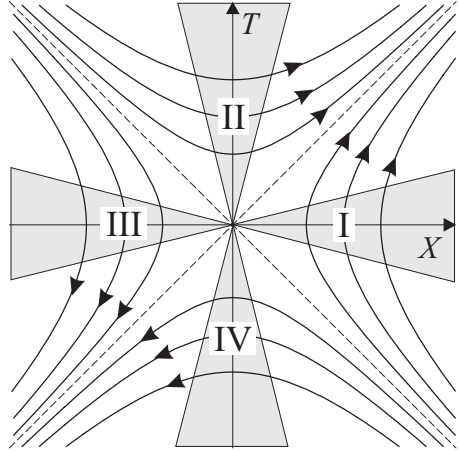}
\hfill {}
\\
\centering \caption{Polar coordinates in the Minkowskian plane. Solid
lines are hyperbolas of constant $R$. Arrows show increasing of the
angle $\Phi$. Shadowed regions denote the fundamental domain of the
transformation group $\Phi\rightarrow\Phi+2\pi\al$. \label{fpolmi}}
\end{figure}

In this section we perform the coordinate transformation which brings
the metric (\ref{emetrs}) into the usual Lorentzian form and shows that
the Carter--Penrose diagram in Fig.\ref{fcpfou} represents the Minkowskian
plane. We consider three-dimensional Minkowskian space $\MR^{1,2}$ with
the Lorentz metric in Cartesian coordinates
\begin{equation}                                             \label{eflthm}
  ds^2=dT^2-dX^2-dY^2.
\end{equation}
We introduce polar coordinates $0<R<\infty, -\infty<\Phi<\infty$ in each
of the four quadrants in the $T,X$ plane (see Fig.\ref{fpolmi})
\begin{equation*}
\begin{split}
  \text{I}:~~&T=~~R\sinh\Phi,
\\
  &X=~~R\cosh\Phi,
\\
  \text{III}:~~&T=-R\sinh\Phi,
\\
  &X=-R\cosh\Phi,
\end{split}
\qquad
\begin{split}
  \text{II}:~~&T=~~R\cosh\Phi,
\\
  &X=~~R\sinh\Phi,
\\
  \text{IV}:~~&T=-R\cosh\Phi,
\\
  &X=-R\sinh\Phi.
\end{split}
\end{equation*}
This transformation is degenerate on two lines $R=0$, and metric on the
$T,X$ plane becomes
\begin{equation}                                          \label{emimep}
\begin{split}
\text{I,III}:\quad dl^2&=-dR^2+R^2d\Phi^2,
\\
\text{II,IV}:\quad dl^2&=dR^2-R^2d\Phi^2.
\end{split}
\end{equation}

Performing the further transformation
\begin{equation*}
\begin{aligned}
&\text{I,III}:\quad & R&=\frac{\sqrt{c^2-2\al^3\s}}{\al^2}, &\quad \s&<\frac{c^2}{2\al^3}
\\[2mm]
&\text{II,IV}:\quad & R&=\frac{\sqrt{2\al^3\s-c^2}}{\al^2}, &\quad \s&>\frac{c^2}{2\al^3}
\\[2mm]
&\text{I--IV}:\quad& \Phi&=\al\vf, &\quad -\infty&<\vf<\infty,
\\
&& Y&=\al t', &-\infty&<t'<\infty,
\end{aligned}
\end{equation*}
we arrive precisely to metric (\ref{emetrs}), (\ref{etwmes}).

Now integrating equations (\ref{egebtz}) for geodesics becomes trivial.
A general solution is
\begin{equation*}
\begin{split}
  T&=t_0+v_0\tau,
\\
  X&=x_0+v_1\tau,
\\
  Y&=y_0+v_2\tau,
\end{split}
\end{equation*}
and depends on six arbitrary constants: a position of a point in the
Minkowskian space $t_0,x_0,y_0$ and a velocity $v_0,v_1,v_2$ of the
geodesic which goes through this point. The inverse transformation, for
example, in the first quadrant is
\begin{equation*}
\begin{split}
  t&=\frac1\al\left[Y+\frac c{\al^2}\arctanh\left(\frac TX\right)\right],
\\
  r&=\frac1\al\sqrt{c^2-\al^4(X^2-T^2)},
\\
  \vf&=\frac1\al\arctanh\left(\frac TX\right).
\end{split}
\end{equation*}
Elementary analysis shows that constants of integration
(\ref{econen})--(\ref{econse}) are
\begin{equation*}
\begin{split}
  C_0&=v_0^2-v_1^2-v_2^2,
\\
  C_1&=-\al v_2,
\\
  C_2&=\frac{cv_2}\al+\al(x_0v_0-t_0v_1).
\end{split}
\end{equation*}

The considered coordinate transformations prove that the Carter--Penrose
diagram in Fig.\ref{fcpfou} represents a Minkowskian plane $\MR^{1,1}$.
The global BTZ solution with zero cosmological constant is thus a
product $\MR\times\MR^{1,1}=\MR^{1,2}$, i.e.\ a flat three-dimensional
Minkowskian space-time.

Now the maximally extended BTZ solution with zero cosmological constant
(\ref{ebtzma}) and infinite range of $\vf$ becomes transparent: it is a
flat three-dimensional Minkowskian space-time $\MR^{1,2}$ without any
singularity and horizons. BTZ solution (\ref{ebtzma}) covers only one
half of it. Maximal extension means that the BTZ solution for $0<\s<\infty$
is prolonged to negative values $-\infty<\s<\infty$. For negative $\s$ we
can define new coordinate $r=\sqrt{2\al\s}$, $\s<0$ and $\al<0$ and again
arrive at the BTZ solution. Thus, two copies of BTZ solution cover the
whole Minkowskian space-time $\MR^{1,2}$ and are smoothly glued together
at $r=0$. The ``singularity'' at $r=0$ is a purely coordinate one which
is clear from the transformation (\ref{echrad}). We stress that the above
analysis was performed for the whole range of the angle $\vf\in\MR$. The
periodicity of the angle will be considered in the next section.
\subsection{Periodicity of $\vf$}
Having in mind that the coordinate $\vf$ is interpreted as the angle
$0<\vf<2\pi$ in the exterior region of the BTZ solution with negative
cosmological constant, we assume that the identification $\vf\sim\vf+2\pi$
takes also place for zero cosmological constant. The transformation
group $\MG:~\vf\rightarrow\vf+2\pi$ acts freely and properly discontinuous
on a line $\vf\in\MR$ but not in the Minkowskian space-time $\MR^{1,2}$.
In the last case it has fixed points and the quotient
space $\MR^{1,2}/\MG$ itself is not a manifold.

Periodicity in $\vf$ means periodicity in the polar angle $\Phi\sim\Phi+2\pi\al$
in the Minkowskian plane $T,X$ considered in the previous section.
The transformation group identifies points along hyperbolas $R=\const$
corresponding to different polar angles $\Phi\sim\Phi+2\pi\al$. Its
action is not defined on two lines $R=0$ where polar coordinates are
degenerate. Considering these lines as the limit of hyperbolas, we
assume that all their points are mapped into the origin under the action
of the transformation group $\MG$. Then the transformation group acting
in the $T,X$ plane has the fundamental domain consisting of four wedges with
identified boundaries shown by shadowed regions in Fig.~\ref{fpolmi}. We
denote them by I-IV according to the number of the corresponding quadrant.
Identifying boundaries of the wedge makes it a cone. The vertexes of cones
are glued in the origin of the Cartesian coordinate system. The
origin is a fixed point for the transformation group and not a point of a
manifold. Indeed, it does not have a neighborhood diffeomorphic to a disc.
At the same time, it can not be excluded from the space-time because it lies
at a finite distance. Thus, we have four cones in the Minkowskian plane
with the common vertex in the origin. They are not conical singularities
because the plane is equipped with the Lorentzian signature metric.

The Killing vector $K_2=\pl_\vf=\al\pl_\Ph$ of the whole three-dimensional
metric (\ref{ebtzma}) is also the Killing vector for two-dimensional metric
(\ref{emimep}). Hence, the transformation group $\MG$ is the isometry for
two-dimensional Minkowskian plane $T,X$.

The identification $\vf\sim\vf+2\pi$ acts also nontrivially on the third
``time'' coordinate $t'$ in (\ref{enewti}), but it remains well defined.

\begin{figure}[h,b,t]
\hfill\includegraphics[height=60mm]{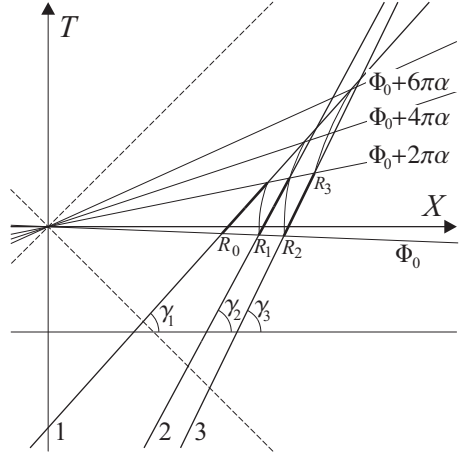}
\hfill {}
\\
\centering \caption{The image of the geodesic $1$ in the fundamental
domain $\Phi_0,\Phi_0+2\pi\al$. It consists of segments of straight lines
with inclinations $\g_1,\g_2,\g_3,\dotsc$ which cross the line $\Phi_0$
respectively at points $R_0,R_1,R_2,\dotsc$.
\label{fangex}}
\end{figure}

Now we analyze the behavior of geodesics in the $T,X$ plane. They are
clearly straight lines before the identification. Let us fix an arbitrary
fundamental domain $(\Phi_0,\Phi_0+2\pi\al)$ and a timelike geodesic $1$,
shown in Fig.~\ref{fangex}. Particle moving along geodesic 1 in the $T,X$
plane wraps around a cone. Its trajectory is represented by the sequence
of segments of straight lines in the fundamental domain, each segment
representing one rotation around the cone. The equation of geodesic 1 in
polar coordinates in the first quadrant is
\begin{equation*}
  R=R_0\frac{\tan\g_1\cosh\Phi_0-\sinh\Phi_0}{\tan\g_1\cosh\Phi-\sinh\Phi},
\end{equation*}
where $\g_1$ is the inclination of the line $1$. Its segment from $R_0$ to
\begin{equation*}
  R_1=R_0\frac{\tan\g_1\cosh\Phi_0-\sinh\Phi_0}
  {\tan\g_1\cosh(\Phi+2\pi\al)-\sinh(\Phi_0+2\pi\al)}
\end{equation*}
belongs to the fundamental domain. The next segment corresponding to
polar angle $\Phi\in(\Phi_0+2\pi\al,\Phi_0+4\pi\al)$ is identified with
the segment of line $2$ lying in the fundamental domain $(\Phi_0,\Phi_0+2\pi\al)$
but having different inclination $\g_2$. Continuing this procedure, we
obtain equation for the $k$-th line
\begin{equation*}
  R=R_{k-1}\frac{\tan\g_k\cosh\Phi_0-\sinh\Phi_0}{\tan\g_k\cosh\Phi-\sinh\Phi},
\end{equation*}
where
\begin{equation}                                        \label{eqangf}
  R_k=R_0\frac{\tan\g_1\cosh\Phi_0-\sinh\Phi_0}
  {\tan\g_1\cosh(\Phi_0+\al_k)-\sinh(\Phi_0+\al_k)},
\end{equation}
and we introduced shorthand notation
\begin{equation*}
  \al_k=2\pi\al k,~~~~k=0,\pm1,\pm2,\dotsc.
\end{equation*}
On the other hand, the segment of line $k$ in the fundamental domain
can be obtained as the projection of the segment of the previous line
$k-1$
\begin{equation}                                        \label{eqrkmo}
  R_k=R_{k-1}\frac{\tan\g_k\cosh\Phi_0-\sinh\Phi_0}
  {\tan\g_k\cosh(\Phi_0+\al_1)-\sinh(\Phi_0+\al_1)}.
\end{equation}
Comparing Eqs.\ (\ref{eqangf}) and (\ref{eqrkmo}), we express the
inclination angle $\g_k$ of the $k$-th line through the original
inclination angle $\g_1$
\begin{equation}                                        \label{eincan}
  \tan\g_k=\frac{\tan\g_1-\tanh\al_{k-1}}
  {1-\tan\g_1\tanh\al_{k-1}}.
\end{equation}
It is important that the inclination angles of segments depend only on
the original inclination angle $\g_1$. They do not depend on the fundamental
domain characterized by the angle $\Phi_0$ and the distance from the
origin $R_0$.

Equation (\ref{eincan}) is also valid for lightlike and spacelike geodesics.

As the consequence of Eq.(\ref{eincan}), we see that segments of all
lightlike geodesics $\tan\g_1=\pm1$ remain lightlike $\tan\g_k=\pm1$.
All segments for timelike and spacelike geodesics are respectively
timelike and spacelike having the same limit
\begin{equation*}
  \lim_{k\to \pm\infty}\tan\g_k=\mp1.
\end{equation*}
We see also that there are no closed timelike geodesics.

\begin{figure}[h,b,t]
\hfill\includegraphics[height=60mm]{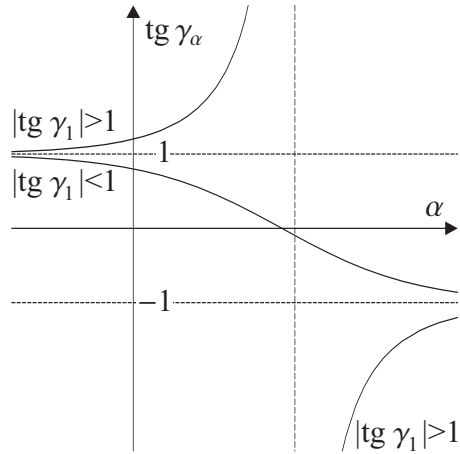}
\hfill {}
\\
\centering \caption{The inclination of the trajectory segment versus continuous
parameter $\al_k\rightarrow\al$. For timelike geodesics $|\tan\g_1|>1$, it has
two branches with the singularity at $\tanh\al\tan\g_1=1$, corresponding to the
turning point. The dependence is smooth for spacelike geodesics $|\tan\g_1|<1$.
\label{ftgalp}}
\end{figure}

To get a deeper insight into the behavior of geodesics we consider
$\al_k$ as a continuous variable
\begin{equation*}
  \tan\g_\al=\frac{\tan\g_1-\tanh\al}
  {1-\tan\g_1\tanh\al}.
\end{equation*}
This function is plotted in Fig.\ref{ftgalp}. For timelike geodesics
$|\tan\g_1|>1$ it is singular when
\begin{equation}                                       \label{edesig}
  \tanh\al\tan\g_1=1
\end{equation}
and has two branches.
The upper and lower branches correspond to the motion of a particle respectively
to the right and left in the first quadrant, the singularity (\ref{edesig})
being the turning point.

All timelike geodesics start and end at the origin of the Cartesian
coordinate system at finite values of proper time parameter making
infinite number of rotations around the cone. Spacelike geodesics which
do not lie entirely in one fundamental region approach the origin on
the left making infinite number of rotations at a finite value of canonical
parameter and go to the right space infinity.

To understand the behavior of geodesics in other quadrants, it is sufficient
to rotate the picture on ninety degrees.

\begin{figure}[h,b,t]
\hfill\includegraphics[height=60mm]{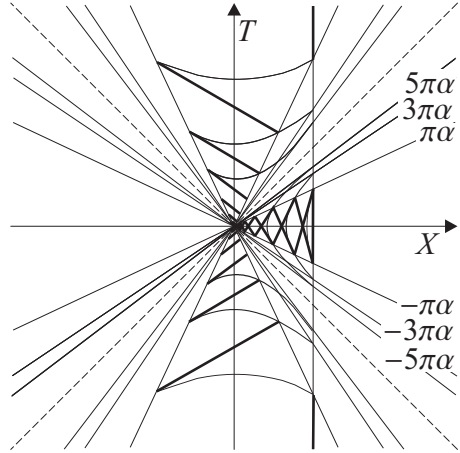}
\hfill {}
\\
\centering \caption{The trajectory of a static particle in the $T,X$ plane.
The fundamental domain is chosen symmetrically $(-\pi,\pi)$ in each quadrant.
After the identification $\vf\sim\vf+2\pi$ it makes infinite number of
rotations around the cone in the fourth quadrant at a finite proper time.
Afterwards it moves to the right in the first quadrant, returns back,
making also an infinite number of rotations near the cone vertex, and
goes to the second quadrant. There, after an infinite number of rotations,
it goes to infinity.
\label{fpares}}
\end{figure}

The behavior of geodesics at the origin of Cartesian coordinates is
not defined. It can be naturally done by unfolding the cones and going
back to Minkowskian plane $T,X$. Then timelike geodesics which do not
cross boundary of the fundamental region IV are continued to the fundamental
region II. Those timelike geodesics which cross the boundary of the
fundamental region IV are continued to the fundamental domain I or III
depending on the inclination. In the region I they move to the right at
a finite distance, then return back to the origin, and are continued to
the fundamental domain II. We draw trajectory of a static particle in
the $T,X$ plane after the identification in Fig.\ref{fpares}.

Timelike curves $R=\const$ (which are not geodesics) are closed loops
after the identification $\Phi\sim\Phi+2\pi\alpha$. Thus, any particle
with constant acceleration (in terms of Cartesian coordinates) has a
closed timelike world line at regions I,III.

\begin{figure}[h,b,t]
\hfill\includegraphics[height=60mm]{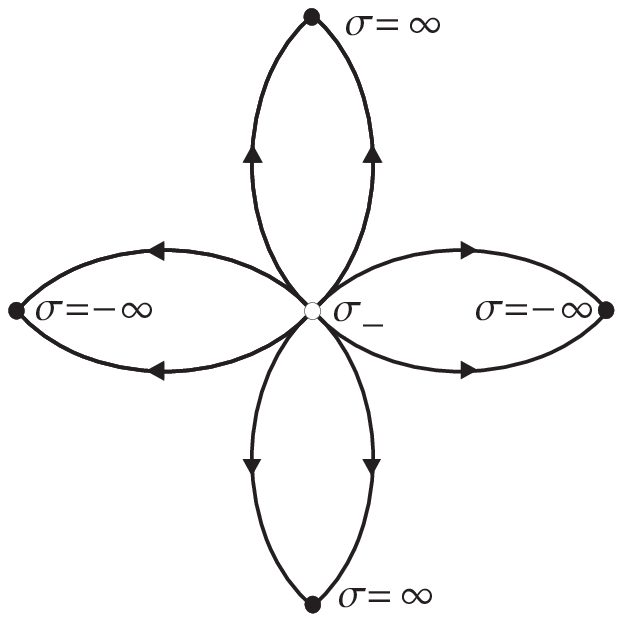}
\hfill {}
\\
\centering \caption{The Carter--Penrose diagram for the Minkowskian plane
after the polar angle identification $\Phi\sim\Phi+2\pi\al$.
\label{flower}}
\end{figure}

To conclude this section, we draw the Carter--Penrose diagram for the
Minkowskian plane after the polar angle identification $\Phi\sim\Phi+2\pi\al$
in Fig.\ref{flower}. It looks like a flower.
\section{Euclidean version of the interior of the BTZ solution}
The Euclidean version of the BTZ solution was considered in \cite{CarTei95}
when analyzing its thermodynamical properties. In this section we
analyze global structure and geodesics of the Euclidean BTZ solution
for zero cosmological constant.

Here, we use many notations from the previous section though their meaning in the
Euclidean case is often different. We hope that this step does not cause
any inconvenience and simplifies the comparison of the two cases.

The Euclidean counterpart of the BTZ metric (\ref{ebtzso}) is given by
the transformation $r\rightarrow ir$, $l\rightarrow il$, and $J\rightarrow-J$.
Then the BTZ metric becomes
\begin{equation}                                        \label{eclbtz}
  ds^2=\left(-\al^2-\frac{c^2}{r^2}+\frac{r^2}{l^2}\right)dt^2
  +\frac{dr^2}{-\al^2-\frac{c^2}{r^2}+\frac{r^2}{l^2}}
  +r^2\left(d\vf-\frac c{r^2}dt\right)^2.
\end{equation}
Let $r_\pm$ denote two positive roots
\begin{equation*}
  r_\pm^2=\frac{\al^2l^2}2\left(\sqrt{1+\frac{4c^2}{\al^4l^2}}\pm1\right).
\end{equation*}
Then metric (\ref{eclbtz}) is degenerate at $r=r_+$, has Euclidean signature
$(+++)$ for $r_+<r<\infty$, and Lorentzian signature $(+--)$ for $0<r<r_+$.
In fact, this metric describes two disjoint spaces: one for $r_+<r<\infty$
with the Euclidean signature metric, and the other for $0<r<r_+$ with
the Lorentzian metric. This phenomena was demonstrated in the two-dimensional
case where connected Lorentzian surface breaks into disconnected
Euclidean pieces along horizons when going to the Euclidean signature
metric \cite{KaKlKu99}, horizons giving rise to possible conical
singularities. The absence of conical singularities is precisely the
definition of the Hawking temperature. The same phenomenon occurs also
for zero cosmological constant (see Sec.~\ref{sgeoeu}).

In the limit $l\rightarrow\infty$, metric (\ref{eclbtz}) reduces to the
metric having Lorentzian signature everywhere and cannot be considered as
the Euclidean version of (\ref{ebtzma}). The Euclidean counterpart of the
BTZ metric for zero cosmological constant (\ref{ebtzma}) is given by the
transformation $t\rightarrow iz$, $\vf\rightarrow i\vf$:
\begin{equation}                                        \label{eclinm}
  ds^2=\al^2 dz^2+\frac{dr^2}{\al^2-\frac{c^2}{r^2}}+r^2d\vf^2-2cd\vf dz.
\end{equation}
The range of coordinates is taken as
\begin{equation}                                        \label{eraecb}
  r\in(0,\infty),~~~~\vf\in(0,2\pi),~~~~z\in(-\infty,\infty).
\end{equation}
This metric is degenerate at the horizon $r_-=c/\al$, has Euclidean
signature $(+++)$ for outer region $r_-<r<\infty$, and Lorentzian signature
$(+--)$ for inner region $0<r<r_-$.

So, our starting point is metric (\ref{eclinm}), (\ref{eraecb}) which is
the Euclidean version of (\ref{ebtzma}). The problem is to find the
space described by this metric. We show below that it describes two
disjoint maximally extended manifolds, each being topologically equivalent
to the Euclidean space $\MR^3$ with the wedge cut out or added to it.
\subsection{Transformation to Cartesian coordinates}
First, we consider the region $c/\al<r<\infty$, where metric (\ref{eclinm})
has Euclidean signature. This domain can be easily shown to be diffeomorphic
to the whole Euclidean space $\MR^3$ with the wedge of angle $2\pi\theta$,
where $\theta=\al-1$, cut out or added. For negative and positive $\theta$
the wedge is respectively cut out or added to $\MR^3$.

Let $X,Y,Z$ be Cartesian coordinates in $\MR^3$
\begin{equation}                                        \label{euccap}
  ds^2=dX^2+dY^2+dZ^2=dR^2+R^2d\Phi^2+dZ^2,
\end{equation}
where $R,\Phi$ are polar coordinates in the $X,Y$ plane. Then
the coordinate transformation:
\begin{equation}                                        \label{ectreu}
\begin{aligned}
  R&=\frac r\al\sqrt{1-\frac{c^2}{\al^2r^2}},&~~~~\frac c\al&<r<\infty,
\\
  \Phi&=\al\vf, & 0&<\Phi<2\pi\al,
\\
  Z&=\al z-\frac c\al \vf, & -\infty&<z<\infty,
\end{aligned}
\end{equation}
brings metric to the form (\ref{eclinm}). The wedge is cut out or added to the
Euclidean space because the angle $\Phi$ ranges from $0$ to $2\pi\al$.
This is determined by the original range of coordinates (\ref{eraecb}).
The sides of the wedge are identified (glued together). The axis $R=0$ is
mapped into the horizon $r_-$ which is now the surface of the cylinder.

The inverse transformation to (\ref{ectreu}) is
\begin{equation*}
\begin{split}
  z&=\frac1\al\left[Z+\frac c{\al^2}\arctan\left(\frac YX\right)\right],
\\
  r&=\frac1\al\sqrt{c^2+\al^4(X^2+Y^2)},
\\
  \vf&=\frac1\al\arctan\left(\frac YX\right).
\end{split}
\end{equation*}

The inner region $0<r<r_-$ is diffeomorphic to the Minkowskian space-time
$\MR^{1,2}$ with the metric
\begin{equation}                                        \label{elobtm}
  ds^2=-dX^2-dY^2+dZ^2=-R^2d\Phi^2-R^2d\Phi^2+dZ^2.
\end{equation}
The coordinate transformation
\begin{equation}                                        \label{ectren}
\begin{aligned}
  R&=\frac r\al\sqrt{\frac{c^2}{\al^2r^2}-1},&~~~~0&<r<\frac c\al,
\\
  \Phi&=\al\vf, & 0&<\Phi<2\pi\al,
\\
  Z&=\al z-\frac c\al \vf, & -\infty&<z<\infty.
\end{aligned}
\end{equation}
transforms metric (\ref{elobtm}) also in the form (\ref{eclinm}). The range
of the angle $\Phi$ differs from $2\pi$. Therefore, the wedge is cut out
or added to the Minkowskian space-time as in the previous case. The axis
$R=0$ and infinity $R=\infty$ are respectively mapped into the horizon
$r_-$ and the axis $r=0$.
\subsection{Geodesics                                 \label{sgeoeu}}
To answer the question what is described by metric (\ref{eclinm}),
(\ref{eraecb}), we must analyze the behavior of geodesics. Equations
for geodesics (\ref{egebtn}) are
\begin{equation}                                        \label{egebez}
\begin{split}
  \ddot z&=-2\frac{cr}{\al^2 r^2-c^2}\dot r\dot \vf,
\\
  \ddot r&=\frac{c^2}{r(\al^2r^2-c^2)}\dot r^2+\frac{\al^2r^2-c^2}r\dot\vf^2,
\\
  \ddot \vf&=-2\frac{\al^2r}{\al^2r^2-c^2}\dot r\dot\vf.
\end{split}
\end{equation}
The only difference from Eqs.(\ref{egebtz}) is the substitution $t\rightarrow z$
and the sign before the second term in the second equation.

There are three conservation laws:
\begin{equation}                                        \label{egeucb}
\begin{split}
  C_0&=\al^2\dot z^2+\frac{r^2}{\al^2r^2-c^2}\dot r^2+r^2\dot\vf^2
  -2c\dot\vf\dot z,
\\
  C_1&=~\al^2\dot z-c\dot\vf,
\\
  C_2&=-c\dot z+r^2\dot\vf.
\end{split}
\end{equation}
The last two conservation laws correspond to two Killing vectors $K_1=\pl_z$
and $K_2=\pl_\vf$.

A general solution of Eqs.(\ref{egebez})
\begin{equation*}
\begin{split}
  X&=x_0+v_1\tau,
\\
  Y&=y_0+v_2\tau,
\\
  Z&=z_0+v_3\tau,
\end{split}
\end{equation*}
depends on six arbitrary constants $(x_0,y_0,z_0)$ and $(v_1,v_2,v_3)$
parameterizing the point and the direction of a geodesic at this point.
Elementary analysis shows that
\begin{equation*}
\begin{split}
  C_0&=v_1^2+v_2^2+v_3^2,
\\
  C_1&=\al v_3,
\\
  C_2&=-\frac c\al v_3+\al(x_0v_2-y_0v_1).
\end{split}
\end{equation*}

To make the difference between Lorentzian and Euclidean cases clear, we
solve Eqs. (\ref{egeucb}) with respect to the first derivatives
\begin{equation}                                        \label{efieod}
\begin{split}
  \dot z&=\frac{r^2C_1+cC_2}{\al^2r^2-c^2},
\\
  \dot r^2&=\frac{\al^2r^2-c^2}{r^2}C_0-\frac{r^2C_1^2+2cC_1C_2+\al^2C_2^2}{r^2},
\\
  \dot\vf&=\frac{cC_1+\al^2C_2}{\al^2r^2-c^2}.
\end{split}
\end{equation}
The essential difference from the Lorentzian case (\ref{efilod}) is the sign
before the second term in the expression for $\dot r^2$. At the horizon
$r_-=c/\al$ we have
\begin{equation*}
  \dot r^2=-\frac{(cC_1+\al^2C_2)^2}{c^2}
\end{equation*}
We see that horizon is reached only by geodesics with $cC_1+\al^2C_2=0$
because the right hand side of this equation must be positive. For these
geodesics
\begin{equation*}
  \dot\vf=0,~~~~\dot z=-\frac{C_2}c.
\end{equation*}
This means that in each plane $z=\const$ only radial geodesics reach the
surface of the cylinder $r=r_-$. Therefore, there are not enough geodesics
to consider the boundary of the cylinder as a two-dimensional surface.
Moreover, the circumference of the cylinder measured with metric
(\ref{eclinm}) is zero. We conclude that the boundary of the cylinder
$r=r_-$ is, in fact, a line. For the Lorentzian signature metric, there is
no restriction on geodesics which reach the horizon due to the difference
in signs.

Previous analysis indicates that the connected Lorentzian manifold breaks
into disconnected pieces along horizons for the Euclidean case. To prove this,
we consider another coordinate transformation $R,\Phi,Z\rightarrow f,\psi,\z$:
\begin{equation}                                        \label{ecotrj}
\begin{split}
  R&=\frac f\al,
\\
  \Phi&=\al\psi,
\\
   Z&=\z-c\psi.
\end{split}
\end{equation}
In these coordinates the flat metric (\ref{euccap}) becomes
\begin{equation}                                        \label{ejteme}
  ds^2=\frac 1{\al^2}df^2+(f^2+c^2)d\psi^2+d\z^2-2cd\z d\psi,
\end{equation}
but now the angle range is $\psi\in(0,2\pi)$, and coordinates $f,\psi,\z$
cover the whole $\MR^3$ and nothing else. (We use the unusual notation
$f$ for the radial coordinate because in the next section it is
transformed $f=f(\rho)$.) This metric describes
the space which is topologically $\MR^3$ with the ``shifted'' conical
singularities along the $\z$ axis. In the case $c=0$, we have ordinary
conical singularity in each section $\z=\const$. This space is
geodesically complete at infinity $f\rightarrow\infty$ because all
geodesics in the original $X,Y,Z$ coordinate are complete. Thus, the
exterior region $r_-<r<\infty$ of the Euclidean BTZ metric for zero
cosmological constant describes Euclidean space with ``shifted''
conical singularities at the $\z$ axis. It is the maximally extended
manifold.

Metric (\ref{ejteme}) is the Euclidean version of the metric considered
in \cite{DeJatH84} and has straightforward interpretation in solid state
physics considered in the next section.

Similar analysis can be performed for the interior region, and we do not
repeat it here. In fact, it is obvious from coordinate transformation
(\ref{ectren}) that the axis $z$ for metric (\ref{eclinm}) represents
infinity and the surface of the cylinder is the ``shifted'' conical
singularity at the $Z$ axis of three-dimensional Minkowskian space-time
(\ref{elobtm}). This manifold is also maximally extended.

Therefore metric (\ref{eclinm}), (\ref{eraecb}) describes two disjoint
maximally extended manifolds: the Euclidean space $\MR^3$ and the
Minkowskian space-time $\MR^{1,2}$ with ``shifted'' conical
singularities at the $Z$ axis.
\section{Solid state physics interpretation           \label{sscwed}}
In the geometric theory of defects (for review see \cite{Katana05})
pure elastic deformations describe diffeomorphisms of the flat Euclidean
space $\MR^3$. The presence of defects: dislocations (defects in elastic media)
and disclinations (defects in the spin structure), gives rise to nontrivial
Riemann--Cartan geometry in $\MR^3$, the curvature and torsion tensor being
interpreted as the surface density of Burgers and Frank vectors, respectively.
In the absence of disclinations, curvature is equal to zero, and we have the
space of absolute parallelism characterized only by nontrivial torsion.
In this case, the $\MS\MO(3)$-connection is a pure gauge defined by the
gauge conditions, and torsion is given by the triad field $e_\mu{}^i$ which
satisfies Euclidean Einstein's equations with nontrivial sources
(energy-momentum tensor). The corresponding Einstein's equations are
written for the induced metric
\begin{equation*}
  g_{\mu\nu}=e_\mu{}^ie_\nu{}^j\dl_{ij},~~~~\dl_{ij}=\diag(+++),
\end{equation*}
while the curvature tensor for the $\MS\MO(3)$-connection remains identically
equal to zero.
For continuous distribution of dislocations, the sources are described by
continuous functions, and for single defects, we have $\dl$-function type
sources in the right hand side of Euclidean Einstein's equations.

Metric (\ref{ejteme}) satisfies free Euclidean Einstein's equations
everywhere except the $\z$ axis, where we have a ``shifted'' conical
singularity. We do not analyze the corresponding source here but
simply give physical interpretation of this solution.

The coordinate transformation from the flat Euclidean space (\ref{ecotrj})
has the following interpretation in solid state physics. At the beginning,
we have undeformed elastic media which is the flat Euclidean space
$\MR^3$. For negative deficit angle $2\pi\theta$, the wedge with
the edge along the $Z$ axis and the angle $2\pi|\theta|$ is cut out from
the media (see Fig.). For positive deficit angle, the wedge of the
same media is added.
\begin{figure}[h,b,t]
\hfill\includegraphics[height=60mm]{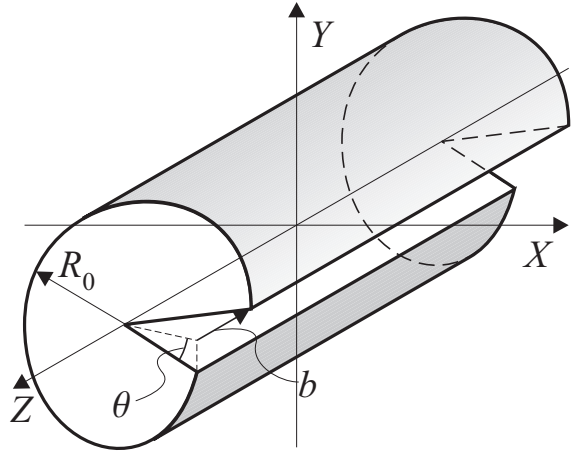}
\hfill {}
\\
\centering \caption{The combined wedge and screw dislocation. The picture
is drawn for negative deficit angle $\theta$ and the Burgers vector $b$
which is antiparallel to the $Z$ axis. \label{fwescd}}
\end{figure}
Then the lower side of the cut is moved along the $Z$ axis in the opposite
direction on the distance $b=2\pi c$, where $b$ is the Burgers vector, and
both sides of the cut are glued together.
The cut out or added wedge of media corresponds to the wedge dislocation,
and the displacement of the lower side of the cut along the $Z$ axis
describes the screw dislocation. Both types of dislocations are experimentally
observed defects in real crystals (see i.e.\ \cite{Amelin64}). Thus,
metric (\ref{ejteme}) qualitatively describes nontrivial geometry around
combined wedge and screw dislocations. For this metric, we can construct
triad field and the corresponding torsion tensor which is equal to the
surface density of the Burgers vector.

We used the word ``qualitatively'' because a solution of Einstein's
equations can be written in an arbitrary coordinate system and does not
depend on Lame coefficients which characterize the elastic properties
of media. The coordinate transformation (\ref{ecotrj}) describes
only cut and paste process of defect creation. In reality, the media
comes to the equilibrium state after gluing both sides of the cut, and
this process is driven by the elasticity theory equations. Therefore,
the induced metric must depend on Lame coefficients. To consolidate
gravity and elasticity theories we proposed to use the elastic gauge
\cite{Katana03}. It is given by the following construction. First,
we fix the cylindrical coordinates $\rho,\psi,\z$ in the Euclidean
space related to the considered problem. The flat metric and the
triad field are marked with a circle over a symbol
\begin{equation*}
\begin{split}
  ds^2&=\overset\circ g{}_{\mu\nu}dx^\mu dx^\nu=d\rho^2+\rho^2d\psi^2+d\z^2,
\\
  \overset\circ e{}_\mu{}^i&=
  \begin{pmatrix} 1 & 0 & 0 \\ 0 & \rho & 0 \\ 0 & 0 & 1 \end{pmatrix}
\end{split}
\end{equation*}
Then the elastic gauge is chosen to be
\begin{equation}                                        \label{elgbtz}
  \overset\circ g{}^{\mu\nu}\overset{\circ}{\nb}_\mu e_{\nu i}
  +\frac\s{1-2\s}\overset\circ e{}^\mu{}_i\overset\circ\nb_\mu e^T=0,
\end{equation}
where a circle over the covariant derivative $\overset\circ \nb{}_\mu$
means that it is constructed for flat metric $\overset\circ g{}_{\mu\nu}$;
$e^T=\overset\circ e{}^\mu{}_ie_\mu{}^i$, and $\s=\const$ is the Poisson
ratio defined by the Lame coefficients \cite{LanLif70}.

Physical meaning of this gauge condition is based on the linear
approximation. Suppose defects are absent, and we have only elastic
deformations described by the displacement vector field $u^i(x)$ which
parameterizes diffeomorphisms of $\MR^3$. For simplicity, we assume
that it is given in the Cartesian coordinate system. Than, for small
relative displacements $|\pl_\mu u^i|\ll 1$, the induced triad field
can be chosen in the form (there is a freedom in local $\MS\MO(3)$ rotations):
\begin{equation}                                        \label{elinar}
  e_{\mu i}\approx\dl_{\mu i}-\frac12(\pl_\mu u_i+\pl_i u_\mu).
\end{equation}
In the linear approximation in Cartesian coordinates, Latin and Greek indices
may be identified. Then the elastic gauge condition (\ref{elgbtz}) reduces
to the usual linear elasticity theory equations for the displacement vector
field \cite{LanLif70}
\begin{equation}                                        \label{eqsepu}
  (1-2\s)\triangle u_i+\pl_i\pl_j u^j=0,
\end{equation}
and the reduction of Eq.~(\ref{elgbtz}) to Eq.~(\ref{eqsepu}) is quite
obvious. Introduction of flat covariant derivatives in the elastic gauge
(\ref{elgbtz}) allows us to use arbitrary coordinate systems.

The triad field corresponding to metric (\ref{ejteme}) is
\begin{equation}                                        \label{etrwsc}
  e_\mu{}^i=\begin{pmatrix}
  \dfrac{\raise-.7ex\hbox{$1$}}{\raise .7ex\hbox{$\al$}} & 0 & 0 \\
  0 & f & -c \\ 0 & 0 & 1 \end{pmatrix}.
\end{equation}
The $f,\psi$ part of the triad corresponds to the symmetrized linear
approximation (\ref{elinar}), and the constant unsymmetrical $\psi$
part does not alter the form of the elastic gauge condition. We change
the radial coordinate $f=f(\rho)$ to rewrite the triad in the elastic
gauge. Then the elastic gauge condition (\ref{elgbtz}) reduces to
the Euler ordinary differential equation
\begin{equation}                                      \label{eulgae}
  \frac {f''}\al\left(1+\frac \s{1-2\s}\right)
  +\frac{f'}\rho\left(\frac1\al+\frac\s{1-2\s}\right)
  -\frac f{\rho^2}\left(1+\frac\s{1-2\s}\right)=0.
\end{equation}
This is the same equation which arises for pure wedge dislocation \cite{Katana03}
(the interested reader can find more details there). To uniquely fix the
solution of this equation, we suppose that the media fills only the cylinder
of finite radius $R_0$ (this is needed to avoid divergences) and impose two
additional gauge conditions
\begin{equation*}
  e_\rho{}^\rho|_{\rho=R_0}=1,~~~~e_\psi{}^\psi|_{\rho=0}=0.
\end{equation*}
The first gauge condition means the absence of external forces on the
surface of the cylinder, and the second one corresponds to the absence
of the angular component of the deformation tensor at the core of dislocation.
Afterwards, the solution of Eq.(\ref{eulgae}) is uniquely defined
\begin{equation*}
  f=\frac\al{\g R_0^{\g-1}}\rho^\g,
\end{equation*}
where
\begin{equation*}
  \g=-\theta B+\sqrt{\theta^2B^2+1+\theta},~~~~B=\frac\s{2(1-\s)}.
\end{equation*}
Thus, the Euclidean version of the BTZ solution for zero cosmological
constant (\ref{ejteme}) in the elastic gauge (\ref{elgbtz}) becomes
\begin{equation}                                        \label{eubtze}
  ds^2=\left(\frac\rho {R_0}\right)^{2(\g-1)}d\rho^2
  +\left(\frac{\al^2}{\g^2}\left(\frac\rho{R_0}\right)^{2(\g-1)}\rho^2
  +c^2\right)d\psi^2+d\z^2-2cd\z d\psi.
\end{equation}
This is the exact solution of the Euclidean Einstein equations written in the
elastic gauge. It describes the induced metric around combined wedge and screw
dislocations and nontrivially depends on the Poisson ratio characterizing the
elastic properties of media. Below, this solution is compared with the result
obtained entirely within the elasticity theory without referring to Einstein
equations.

The wedge and screw dislocations are relatively simple linear defects in
solids, and the corresponding displacement vector field can be explicitly
found as the solution to the linear elasticity field equations (\ref{eqsepu}).
The results are well known, and we skip their derivation. The displacement
vector field for the wedge dislocation in cylindrical coordinates
$R,\Phi,Z$ is \cite{Kosevi81}
\begin{equation}                                        \label{eweufi}
  u_R^{(\text{wedge})}=-\theta\frac{1-2\s}{2(1-\s)}\rho\ln\frac\rho{\ex R_0},~~~~
  u_\Psi^{(\text{wedge})}=-\theta\rho\phi,~~~~
  u_Z^{(\text{wedge})}=0,
\end{equation}
where $\ex$ is the base of the natural logarithm. The displacement vector
field for the screw dislocation is \cite{LanLif70}
\begin{equation}                                        \label{esc}
  u_R^{(\text{screw})}=u_\Phi^{(\text{screw})}=0,~~~~
  u_Z^{(\text{screw})}=c\phi=\frac b{2\pi}\phi,
\end{equation}
where $b$ is the Burgers vector. The displacements vectors can be added
in the linear elasticity theory, and the total displacement vector for
combined wedge and screw dislocations becomes
\begin{equation*}
  u=u^{(\text{wedge})}+u^{(\text{screw})}.
\end{equation*}
In our notations, the coordinate transformation corresponding to this
displacement vector field is
\begin{equation*}
  R=\rho-u_R,~~~~\Phi=\phi-\frac1\rho u_\Phi,~~~~Z=\z-u_Z.
\end{equation*}
The induced metric within the elasticity theory becomes
\begin{equation}                                          \label{elmets}
\begin{split}
  ds^2_{\text{(elastic)}}&=dR^2+R^2d\Phi^2+dZ^2=
\\
  &=\left(1+\theta\frac{1-2\s}{1-\s}\ln\frac\rho{R_0}\right)d\rho^2
  +\left[\rho^2\left(1+\theta\frac{1-2\s}{1-\s}\ln\frac\rho{R_0}+\theta\frac1{1-\s}
  \right)+c^2\right]d\phi^2
\\
  &+d\z^2-2cd\z d\phi.
\end{split}
\end{equation}
The linear elasticity theory equations are valid for small relative displacements
$\pl_\mu u^i\ll1$. Therefore, the induced metric in the elasticity theory
is expected to give correct answer for small deficit angle $\theta\ll1$,
small Burgers vector $b/R_0\ll1$, and near the surface of the cylinder
$\rho\sim R_0$.

To compare metrics (\ref{eubtze}) and (\ref{elmets}), it is enough to consider
small deficit angles. For $\theta\ll1$,
\begin{equation*}
  \g\approx1+\theta\frac{1-2\s}{2(1-\s)}.
\end{equation*}
Expanding expression (\ref{eubtze}) in $\theta$, we obtain metric (\ref{elmets})
in the linear approximation.

So, the elasticity theory induced metric reproduces only the linear
approximation of the exact solution of the Euclidean Einstein's equations
within the geometric theory of defects. Metric (\ref{eubtze}) obtained
within the geometric approach is simpler in its form, valid for the whole
range of radius $0<\rho<R_0$, all deficit angles $-1<\theta<\infty$, and
Burgers vectors $b$. The components of the induced metric are proportional
to the stress tensor of media. Therefore, the result obtained within the
geometric theory of defects can be verified experimentally.
\section{Conclusion}
We considered the BTZ black hole solution for zero cosmological constant
in detail. This case describes the interior region of the BTZ black hole
and is simple enough to perform all calculation explicitly. We showed
that points at $r=0$ are just coordinate singularities, and all geometric
quantities are regular there, while the line $r=r_-$ corresponding to the
inner horizon is singular: these points are not points of a manifold.
There are four cones at each point of the line $r=r_-$. The singularity
arises only after the identification of the polar angle $\vf\sim\vf+2\pi$.

The singularity structure is probably the same in a general case for negative
cosmological constant, though it is analyzed only for zero cosmological
constant.

In the Euclidean case, the BTZ solution for zero cosmological constant
breaks into two disjoint manifolds along the horizon $r=r_-$ with
Euclidean and Lorentzian signature metrics. The manifold with the
Euclidean signature metric has straightforward physical interpretation
in the geometric theory of defects describing combined wedge and screw
dislocations in crystals. We showed that the induced metric obtained
entirely within the ordinary elasticity theory provides only the linear
approximation for the exact solution of Einstein's equations. The
Euclidean metric in the elastic gauge obtained from the BTZ solution
depends nontrivially on Lame coefficients and can be measured
experimentally.

The Euclidean version of the BTZ solution for negative cosmological
constant has also straightforward interpretation in the geometric theory
of defects, but in this case it describes continuous distribution of
dislocations and is not so visual.

{\bf Acknowledgments.} We are grateful to I.~L.~Shapiro for discussions.
One of the authors (M.K.) thanks the Universidade Federal de Juiz de Fora
for the hospitality, the FAPEMIG, the Russian Foundation of Basic Research
(Grant No.\ 05-01-00884), and the Program for Supporting Leading
Scientific Schools (Grant No.\ NSh-6705.2006.1) for financial support.


\begin{thebibliography}{10}

\bibitem{BaTeZa92}
M.~Ba\~nados, C.~Teitelboim, and J.~Zanelli.
\newblock Black hole in three-dimensional spacetime.
\newblock {\em Phys.\ Rev.\ Lett.}, 69(13):1849--1851, 1992.

\bibitem{AyMaZa04}
E.~Ayon-Beato, C.~Martinez, and J.~Zanelli.
\newblock {\em Phys.\ Rev.} D70, 044027, 2004.

\bibitem{Carlip98}
S.~Carlip.
\newblock {\em Quantum Gravity in $2+1$ Dimensions}.
\newblock Cambridge University Press, Cambridge, 1998.

\bibitem{BaHeTeZa93}
M.~Ba\~nados, M.~Henneaux, C.~Teitelboim, and J.~Zanelli.
\newblock Geometry of the $2+1$ black hole.
\newblock {\em Phys.\ Rev.}, D48(4):1506--1525, 1993.

\bibitem{Steif96}
A.~R. Steif.
\newblock Supergeometry of three-dimensional black holes.
\newblock {\em Phys.\ Rev.\ D}, 53(10):5521--5526, 1996.

\bibitem{KatVol92}
M.~O. Katanaev and I.~V. Volovich.
\newblock Theory of defects in solids and three-dimensional gravity.
\newblock {\em Ann.\ Phys.}, 216(1):1--28, 1992.

\bibitem{KatVol99}
M.~O. Katanaev and I.~V. Volovich.
\newblock Scattering on dislocations and cosmic strings in the geometric theory
  of defects.
\newblock {\em Ann.\ Phys.}, 271:203--232, 1999.

\bibitem{Katana03}
M.~O. Katanaev.
\newblock Wedge dislocation in the geometric theory of defects.
\newblock {\em Theor.\ Math.\ Phys.}, 135(2):733--744, 2003.

\bibitem{Katana04}
M.~O. Katanaev.
\newblock One-dimensional topologically nontrivial solutions in the {S}kyrme
  model.
\newblock {\em Theor.\ Math.\ Phys.}, 138(2):163--176, 2004.

\bibitem{Katana05}
M.~O. Katanaev.
\newblock Geometric theory of defects.
\newblock {\em Physics -- Uspekhi}, 48(7):675--701, 2005.

\bibitem{Wolf72}
J.~A. Wolf.
\newblock {\em Spaces of constant curvature}.
\newblock University of California, Berkley, California, 1972.

\bibitem{Starus63}
A.~Staruszkiewicz.
\newblock Gravitational theory in three-dimensional space.
\newblock {\em Acta Phys.\ Polon.}, 24(6(12)):735--740, 1963.

\bibitem{Clemen76}
G.~Clem\'ent.
\newblock Field--theoretic particles in two space dimensions.
\newblock {\em Nucl.\ Phys.}, B114:437--448, 1976.

\bibitem{DeJatH84}
S.~Deser, R.~Jackiw, and G.~'t~Hooft.
\newblock Three-dimensional {E}instein gravity: Dynamics of flat space.
\newblock {\em Ann.\ Phys.}, 152(1):220--235, 1984.

\bibitem{Katana00A}
M.~O. Katanaev.
\newblock Global solutions in gravity. {L}orentzian signature.
\newblock {\em Proc.\ Steklov Inst.\ Math.}, 228:158--183, 2000.

\bibitem{FaGaSe93}
C.~Farina, J.~Gamboa, and A.~J. Segu\'i-Santonja.
\newblock Motion and trajectories of particles around three-dimensional black
  holes.
\newblock {\em Class. Quantum Grav.}, 10(11):L193--L200, 1993.

\bibitem{CrMaPe94}
N.~Cruz, C.~Mart\'inez, and L.~Pe\~na.
\newblock Geodesic structure of the $2+1$ black hole.
\newblock {\em Class. Quantum Grav.}, 11(11):2731--2740, 1994.

\bibitem{CarTei95}
S.~Carlip and C.~Teitelboim.
\newblock Aspects of black hole quantum mechanics and thermodynamics in 2+1
  dimensions.
\newblock {\em Phys.\ Rev.\ D}, 51(2):622--631, 1995.

\bibitem{KaKlKu99}
M.~O. Katanaev, T.~Kl\"osch, and W.~Kummer.
\newblock Global properties of warped solutions in general relativity.
\newblock {\em Ann.\ Phys.}, 276:191--222, 1999.

\bibitem{Amelin64}
S.~Amelinckx.
\newblock {\em The direct observation of dislocations}.
\newblock Academic Press, New York and London, 1964.

\bibitem{Kosevi81}
A.~M. Kosevich.
\newblock {\em Physical mechanics of real crystals}.
\newblock Naukova dumka, Kiev, 1981.
\newblock [in Russian].

\bibitem{LanLif70}
L.~D. Landau and E.~M. Lifshits.
\newblock {\em Theory of Elasticity}.
\newblock Pergamon, Oxford, 1970.

\end{thebibliography}
\end{document}